\documentclass[a4paper,fleqn]{cas-sc}

\usepackage[numbers,sort&compress]{natbib}

\usepackage{tabularx}
\usepackage{listings}

\newcolumntype{P}[1]{>{\raggedright\arraybackslash}p{#1}}
\newcolumntype{Q}{>{\raggedright\arraybackslash}X}
\newcommand{\MateraTableSetup}{\small\setlength{\tabcolsep}{6pt}\renewcommand{\arraystretch}{1.2}}
\newcommand{\MateraTH}[1]{\textbf{#1}}

\definecolor{codegreen}{rgb}{0,0.6,0}
\definecolor{codegray}{rgb}{0.5,0.5,0.5}
\definecolor{codepurple}{rgb}{0.2235,0.1608,0.4941}
\definecolor{backcolour}{rgb}{0.98,0.98,0.98}

\lstdefinelanguage{json}{
    morestring=[b]",
    morestring=[d]',
    string=[s]{"}{"},
    comment=[l]{:\ "},
    morecomment=[l]{:"},
}

\lstdefinestyle{mystyle}{
    backgroundcolor=\color{backcolour},
    commentstyle=\color{codegreen},
    keywordstyle=\color{magenta},
    numberstyle=\tiny\color{codegray},
    stringstyle=\color{codepurple},
    basicstyle=\scriptsize\ttfamily,
    breaklines=true,
    captionpos=b,
    keepspaces=true,
    numbers=left,
    numbersep=2pt,
    showstringspaces=false,
    tabsize=2
}
\def\tsc#1{\csdef{#1}{\textsc{\lowercase{#1}}\xspace}}
\tsc{WGM}
\tsc{QE}
\begin{document}

\let\WriteBookmarks\relax
\def\floatpagepagefraction{1}
\def\textpagefraction{.001}

\title [mode = title]{M-CODE: Materials Categorization via Ontology, \newline Dimensionality and Evolution}  

\author[1]{Vsevolod Biryukov}%[<options>]
\orcidauthor{0009-0003-4582-9509}{Vsevolod Biryukov}
\affiliation[1]{organization={Exabyte Inc. (Mat3ra.com)},
%            addressline={}, 
            city={Walnut Creek},
%          citysep={}, % Uncomment if no comma needed between city and postcode
            postcode={94596}, 
            state={CA},
            country={USA}}

\author[2,3,4]{Kamal Choudhary}%[]
\orcidauthor{0000-0001-9737-8074}{Kamal Choudhary}

\affiliation[2]{organization={Department of Materials Science and Engineering, Johns Hopkins University},
%            addressline={}, 
            city={Baltimore},
%          citysep={}, % Uncomment if no comma needed between city and postcode
            postcode={21218}, 
            state={MD},
            country={USA}}
\affiliation[3]{organization={Department of Electrical and Computer Engineering, Johns Hopkins University},
%            addressline={}, 
            city={Baltimore},
%          citysep={}, % Uncomment if no comma needed between city and postcode
            postcode={21218},
            state={MD},
            country={USA}}
\affiliation[4]{organization={DeepMaterials LLC (deepmaterials.org)},
%            addressline={}, 
            city={Silver Spring},
%          citysep={}, % Uncomment if no comma needed between city and postcode
            postcode={20906}, 
            state={MD},
            country={USA}}

\author[1]{Timur Bazhirov}%[]
\orcidauthor{0000-0003-3719-522X}{Timur Bazhirov}

\shortauthors{Biryukov et.al.}
\shorttitle{M-CODE: Materials Categorization via Ontology, Dimensionality and Evolution}

% Corresponding author text
\cortext[1]{Corresponding author: timur@mat3ra.com}


\begin{abstract}
  The rapid advancement of artificial intelligence in materials science requires data standards and data management practices that can capture the complexity of real-world structures, including surfaces, interfaces, defects, and dimensionality reduction. We present M-CODE - Materials Categorization via Ontology, Dimensionality and Evolution - a compact categorization system that links materials-science-specific terminology to a set of reusable concepts as building blocks and provenance-aware transformations. M-CODE classifies structures by dimensionality, structural complexity (from pristine to compound pristine, defective, and processed), and variants that capture common structure creation and evolution approaches. A practical implementation of the categorization is provided in an open-source codebase that includes JSON schemas, examples, and Python and TypeScript types/interfaces, designed to support reproducible dataset generation, validation, and community contributions.
\end{abstract}

\maketitle

% Main text
\section{Introduction}
\label{sec:introduction}

    Modern materials research increasingly relies on digital data and automation to shorten the time from hypothesis to validated insight. Large-scale open databases and workflow management systems have enabled high-throughput computational screening and accelerated discovery across broad application domains \cite{jain2013materialsproject, curtarolo2012aflowlib, saal2013openQMD, pizzi2016aiida, nomad, choudhary2020joint}. At the same time, many machine-learning datasets and benchmarks remain dominated by idealized three-dimensional crystals, while real-world performance often depends on surfaces, reconstructions, disorder, and defects, as well as heterogeneous interfaces that determine transport, catalysis, and stability \cite{ward2016ml-framework, isayev2017ml-descriptors}.
    
    Bridging this gap requires not only structure generation tooling but also a shared language for describing what structures are, how they were constructed, and how to reproduce them. Terminology such as monolayer, slab, interface, or vacancy can be used inconsistently across subfields, while code implementations often re-encode the same concepts in incompatible ways. The resulting fragmentation makes datasets harder to curate, models harder to compare, and provenance harder to track.
    
    M-CODE (Materials Categorization and Ontology via Dimensionality and Evolution) addresses this need with a compact categorization system that links domain-science concepts to reusable software entities and operations. The approach is implemented as open JSON schemas and examples and is designed to support validation, exchange, and provenance-aware regeneration of realistic structures. The work builds on our experience at Mat3ra.com \cite{mat3ra-com-website} developing an open ecosystem for computational materials R\&D since 2015 \cite{EB-ARX-02-2019} and complements established Python tooling for materials analysis and atomistic modeling \cite{ong2013python, larsen2017atomic}.

    The scope of M-CODE is the standardization of structure classes, their construction inputs, and provenance-aware build descriptions for atomistic materials models. M-CODE does not standardize electronic-structure simulation inputs or define a universal property ontology; instead, it targets the intermediate layer that connects realistic structures to reusable build operations.

    The main contributions of this work are:
        (1) a compact categorization of realistic structures by domain, dimensionality, category, and variants, with stable tags for common classes;
        (2) an implementation-oriented ontology of entities and operations that can be mapped to JSON schemas and to corresponding classes and methods;
        (3) permissive, extensible metadata conventions for recording build provenance in generated materials;
        (4) open schema and example artifacts distributed with a reference implementation \cite{exabyteESSEGithubRepo, mat3ra-esse-pypi-package}.
    
    The remainder of this manuscript is organized as follows. Section~\ref{sec:methodology} introduces the M-CODE framework and the associated ontology of entities and operations. Section~\ref{sec:results} presents the categorization tables and highlights the schema artifacts and access points for reuse. Section~\ref{sec:discussion} discusses implications, limitations, and opportunities for community extension. Section~\ref{sec:conclusion} concludes with a brief summary.

\section{Methodology}
\label{sec:methodology}
    
    \subsection{Overview}
    \label{subsec:methodology-general}
    
        M-CODE is designed to bridge a common gap in data-driven materials modeling: a mismatch between idealized training structures and the realistic low-dimensional, defective, and heterogeneous structures that dominate device-relevant behavior. The methodology combines (i) a compact categorization scheme for target structures and (ii) a software-oriented ontology of entities and operations used to build those structures reproducibly.
        
        We use JavaScript Object Notation (JSON) and JSON Schema as a foundation for data organization, validation, and exchange \cite{JSONSchemaDotOrg}.
        In ESSE, the JSON schemas act as a single source of truth: language bindings such as Python and TypeScript interfaces are generated automatically from the schemas to keep definitions consistent across implementations.
        The reference implementation follows object-oriented design to keep concepts modular and reusable. The corresponding codebase is distributed as \texttt{mat3ra-esse}~\cite{mat3ra-esse-pypi-package} PyPi and \texttt{@mat3ra/esse}~\cite{mat3ra-esse-npm-package} NPM packages.

    \subsection{Main Concepts}
    \label{subsec:main-concepts}

        We represent each structure through a combination of its building blocks - simpler structures and a process that combines these building blocks together in a certain way to create the structure in question.
    
        A key choice is to represent materials generation as explicit transformations with provenance. Each structure is created from a \textbf{Configuration} that specifies input structures and physical parameters, consumed by a \textbf{Builder} that returns the resulting \textbf{Material} together with metadata describing the applied transformation.
    
        \refstepcounter{figure}
        \addcontentsline{lof}{figure}{\protect\numberline{\thefigure}{Core concepts of provenance-aware materials generation: configuration, builder, and resulting material.}}
        \begin{center}
            \includegraphics[width=0.99\linewidth,keepaspectratio]{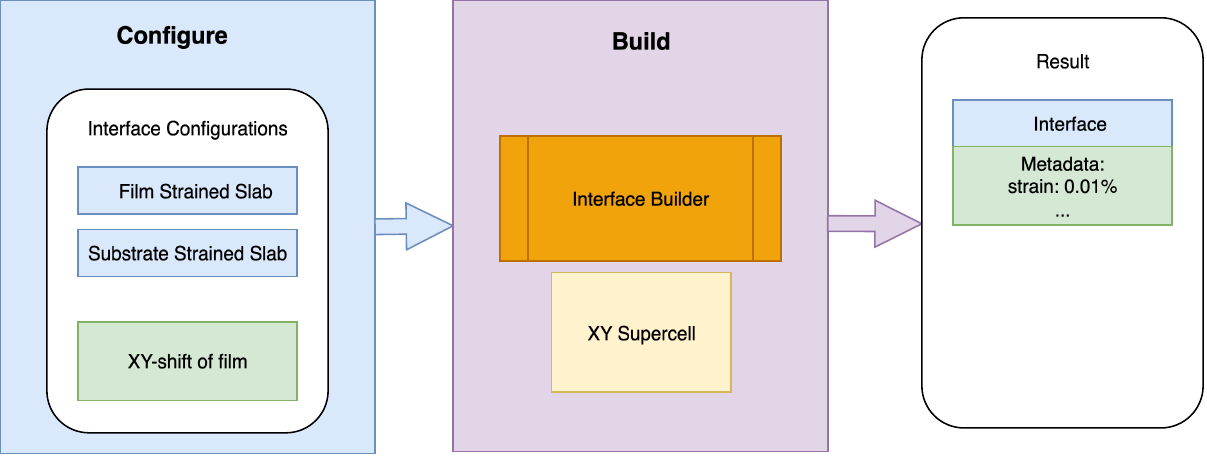}
        \end{center}
        \noindent\textbf{Figure~\thefigure.} Visual representation of the Main concepts of the approach to constructing materials structures. A particular example of creating an interface is depicted. "Configuration", on the left, represents the input - physical components required for creating the target structure. "Build", in the middle, represents the process of combining the input components and specific methods, operations, and parameters related. Finally, "Result", on the right, represents the outcome - resulting material.\label{fig:configure-build-result}

    \subsection{Entities}
    \label{subsec:methodology-entities}

        To connect domain-science descriptions to executable workflows, the data convention organizes materials structures through a composition of entities and operations. Entities are grouped into four categories based on their role in construction.
        The entities and operations are specified as JSON Schemas for validation and interchange, and are realized in code as corresponding classes and methods. This makes it possible to translate a scientific description of a structure into a machine-readable configuration that can be built reproducibly.

        For example, a ``Ni(001) slab with $N$ layers and $10$~\AA{} vacuum'' can be encoded as a slab configuration constructed from a \texttt{crystal} entity (Ni bulk), auxiliary descriptors such as \texttt{miller\_indices} and termination, and a \texttt{vacuum} entity. A slab builder then applies the required operations (e.g., plane selection, layer stacking, and combination with vacuum) and records the choices and parameters as provenance metadata in the resulting material.

        \textbf{Core entities} represent irreducible physical pieces that must exist somewhere in every model (Table~\ref{tab:categorization-software-core-entities}). Examples include a bulk \texttt{crystal} (lattice + basis), \texttt{vacuum} (to break periodicity), a stand-alone \texttt{atom} (e.g., for adatoms or interstitials), and a \texttt{crystal\_site} used as a reference point inside a parent crystal.

        \textbf{Auxiliary entities} are descriptors that parameterize builds and transformations without adding atoms themselves (Table~\ref{tab:categorization-software-aux-entities}). Examples include \texttt{supercell\_matrix\_3d} (replication), \texttt{miller\_indices} and \texttt{termination} (surface orientation and stoichiometry), and a \texttt{shape\_function} used to carve cutouts or finite shapes.

        \textbf{Reusable entities} encode validated macro blocks that are built once and referenced across workflows (Table~\ref{tab:categorization-software-reusable-entities}). Examples include \texttt{supercell}, strain-applied crystals, oriented plane stacks and unique atomic layers, a \texttt{slab\_unit\_cell}, and low-dimensional building blocks such as \texttt{ribbon\_section} and \texttt{nanoparticle}.

        \begin{table*}[!htbp]
            \centering
            \MateraTableSetup
            \begin{tabularx}{\textwidth}{|P{1.2cm}|P{2.2cm}|Q|Q|}
                \hline
                \MateraTH{Dim.} & \MateraTH{Type} & \MateraTH{Purpose} & \MateraTH{Required / optional fields} \\
                \hline
                3D & crystal & Ideal bulk lattice + basis. & (none -- attributes live in the enclosing JSON) \\
                \hline
                3D & void & Delete all atoms inside a shape. & center\_coordinate, shape (sphere, box, custom f(x,y,z)) \\
                \hline
                2D & vacuum & Break periodicity, lower dimensionality. & direction, size (\AA) \\
                \hline
                0D & crystal\_site & Reference point inside a parent crystal. & coordinate, id \\
                \hline
                0D & atom & Stand-alone atom (used for adatoms, interstitials). & chemical\_element \\
                \hline
            \end{tabularx}
            \caption{The list of Core Entities. Columns: \textbf{Dim.}-dimensionality context (3D, 2D, 0D) where the entity is used; \textbf{Type}-entity name; \textbf{Purpose}-what the entity represents; \textbf{Required / optional fields}-schema fields needed to define the entity.}
            \label{tab:categorization-software-core-entities}
        \end{table*}
    
        \begin{table*}[!htbp]
            \centering
            \MateraTableSetup
            \begin{tabularx}{\textwidth}{|P{1.2cm}|P{3.2cm}|Q|Q|}
                \hline
                \MateraTH{Dim.} & \MateraTH{Type} & \MateraTH{Purpose} & \MateraTH{Example value} \\
                \hline
                3D & supercell\_matrix\_3d & Repeat-count for unit-cell replication. & [[2,0,0],[0,2,0],[0,0,2]] \\
                \hline
                2D & Miller\_indices, termination & Plane orientation + surface stoichiometry. & [1, -1] ``Si-P6/mmm2'' \\
                \hline
                0D & shape\_function & Formula to carve custom 3D shape. & f(x,y,z) = x\textasciicircum2+y\textasciicircum2-r\textasciicircum2 \\
                \hline
            \end{tabularx}
            \caption{The list of some auxiliary entities. Columns: \textbf{Dim.}-dimensionality context (3D, 2D, 0D) where the entity is used; \textbf{Type}-entity name; \textbf{Purpose}-what the entity describes; \textbf{Example value}-sample data illustrating the entity format.}
            \label{tab:categorization-software-aux-entities}
        \end{table*}
    
        \begin{table*}[!htbp]
            \label{tab:categorization-software-reusable-entities}
            \centering
            \MateraTableSetup
            \begin{tabularx}{\textwidth}{|P{1.2cm}|P{3.2cm}|Q|Q|}
                \hline
                \MateraTH{Dim.} & \MateraTH{Type} & \MateraTH{Key fields} & \MateraTH{The result} \\
                \hline
                3D & supercell & crystal, matrix & Periodic repeat in all axes. \\
                \hline
                3D & strained\_uniform & crystal, strain\_percentage & Isotropic expansion. \\
                \hline
                3D & strained\_non-uniform & crystal, strain\_matrix & Full 3 $\times$ 3 strain. \\
                \hline
                2D & \path{crystal_lattice_planes} & crystal, miller\_indices, use\_conventional\_cell & Infinite stack of (hkl) planes. \\
                \hline
                2D & \path{atomic_layers_unique} & (same as above) + terminations & Unique atomic layers for that orientation. \\
                \hline
                2D & \path{atomic_layers_unique_repeated} & (same as above) + termination\_top, number\_of\_repetitions & N repetitions of those layers. \\
                \hline
                2D & slab\_unit\_cell & stack\_components=[\newline atomic\_layers, vacuum], direction & Minimal (hkl) slab unit cell. \\
                \hline
                1D & ribbon\_section & monolayer, miller\_indices\_2d, width & Generic nanoribbon. \\
                \hline
                1D & ribbon\_section\_hex & monolayer, edge\_type, width, length & Zig-zag / armchair graphene ribbon. \\
                \hline
                1D & wire\_section & shape, orientation & Arbitrary nanowire cut. \\
                \hline
                0D & nanoparticle & supercell, shape, orientation & Finite particle with vacuum. \\
                \hline
            \end{tabularx}
            \caption{The list of some reusable entities. Columns: \textbf{Dim.}-dimensionality of the resulting structure (3D, 2D, 1D, 0D); \textbf{Type}-entity name; \textbf{Key fields}-main input fields referencing other entities; \textbf{The result}-description of what the entity produces.}
        \end{table*}
    
    \subsection{Operations}
    \label{subsec:methodology-operations}

        \textbf{Operations} are algorithms that modify or combine entities.
        We group operations into \textbf{modifications} (transform one entity) and \textbf{combinations} (join multiple entities) (Table~\ref{tab:categorization-software-operations}).
        
        \textbf{Modifications} transform a single entity:

        \begin{itemize}
            \item \textbf{\texttt{strain}}: applies a strain matrix to the lattice, updating atom coordinates consistently.
            \item \textbf{\texttt{repeat}}: replicates a unit cell by an integer matrix to form a larger periodic structure.
            \item \textbf{\texttt{perturb}}: displaces atoms according to a provided function, optionally enforcing isometric constraints.
        \end{itemize}

        \textbf{Combinations} join multiple entities:

        \begin{itemize}
            \item \textbf{\texttt{stack}}: stacks components along a specified direction (e.g., slabs with optional vacuum) to construct composite structures such as interfaces.
            \item \textbf{\texttt{merge}}: merges two structures into a single structure according to a chosen merge method.
        \end{itemize}

        \begin{table*}[!htbp]
            \label{tab:categorization-software-operations}
            \centering
            \MateraTableSetup
            \begin{tabularx}{\textwidth}{|P{2.4cm}|P{2.2cm}|Q|Q|}
                \hline
                \MateraTH{Type} & \MateraTH{Subtype} & \MateraTH{What it does} & \MateraTH{Key parameters} \\
                \hline
                \multirow{3}{*}{Modifications} & strain & Lattice deformation. & matrix \\
                \cline{2-4}
                & repeat & Cartesian replication of the unit cell. & integer\_matrix \\
                \cline{2-4}
                & perturb & Displace atoms via function. & function, is\_isometric \\
                \hline
                \multirow{2}{*}{Combinations} & stack & Stacking of the two materials in a specified direction. & stack\_components, direction \\
                \cline{2-4}
                & merge & Boolean union of two structures. & merge\_structures, method \\
                \hline
            \end{tabularx}
            \caption{The list of operations used to modify or combine entities. Columns: \textbf{Type}-operation category (Modifications change entities in place, Combinations join multiple entities); \textbf{Subtype}-specific operation name; \textbf{What it does}-operation description; \textbf{Key parameters}-main input parameters.}
        \end{table*}

        In M-CODE, \textbf{Evolution} refers to how a target structure is derived from simpler building blocks through a sequence of operations, with each step adding physical complexity. The categorization domains reflect common evolution paths: \textbf{pristine} structures are derived from a single parent material by cleaving and assembling building blocks without introducing new chemistry (e.g., a slab is obtained from a \texttt{crystal} by selecting \texttt{miller\_indices}, forming atomic layers, repeating them to the desired thickness, and stacking with \texttt{vacuum}); \textbf{compound pristine} structures start from \textbf{two or more} pristine parent materials and evolve by \textbf{combinations} such as \texttt{stack} and \texttt{merge} (e.g., interfaces and heterostacks, where slabs may be strained or repeated as geometric alignment steps prior to stacking); \textbf{defective} structures evolve by inserting, removing, or replacing atoms at identified sites (e.g., vacancy, substitution, interstitial, adatom), using auxiliary descriptors such as \texttt{crystal\_site} to define the location; and \textbf{processed} structures represent physical or chemical processing beyond geometric assembly, such as thermal disorder introduced by annealing or chemical functionalization via passivation with atoms. This framing ties each scientific category to an explicit builder workflow and makes the structural class recoverable from the sequence of entities and operations recorded in metadata.

        A typical evolution path can be expressed explicitly as a build graph. For a slab derived from an ideal bulk crystal, one may start from \texttt{crystal}, specify surface orientation via \texttt{miller\_indices}, derive \texttt{crystal\_lattice\_planes}, enumerate \texttt{atomic\_layers\_unique}, and apply \texttt{repeat} to reach the desired thickness (\texttt{atomic\_layers\_unique\_repeated}). The final slab is produced by \texttt{stack} of those layers with a \texttt{vacuum} entity along the chosen direction, yielding a \texttt{slab\_unit\_cell} and then the resulting \textbf{Material}.

        Point defects are created by applying \texttt{merge} operation to combine a parent crystal with a structure containing a site-specific modification. A \textbf{vacancy} is formed by applying \texttt{merge} to the parent with a structure that has a \texttt{void} at the target \texttt{crystal\_site}. An \textbf{interstitial} is formed by applying \texttt{merge} to the parent with a structure that has an \texttt{atom} at the chosen \texttt{crystal\_site}. A \textbf{substitution} combines both operations: a \texttt{void} and a different \texttt{atom} at the same \texttt{crystal\_site} are applied to the parent via \texttt{merge} to replace the original atom. The configuration specifies the parent material, the target site, and the modification type.

    \subsection{Resulting Categorization framework}
    \label{subsec:methodology-categorization-framework}
    
        M-CODE classifies structures along several dimensions: 
        \begin{itemize}
            \item domain (pristine, compound pristine, defective, processed) - based on the Evolution of the structure/entities with increased complexity
            \item dimensionality (3D, 2D, 1D, 0D)
            \item structure category (e.g., slab, interface, vacancy, grain boundary)
            \item optional variants that capture common construction choices (e.g., simple vs reconstructed slab; ZSL vs twisted interface)
        \end{itemize}

        The purpose of the categorization is to describe the \textbf{target material structures} that we want to generate, store, and reuse in datasets and workflows. We maintain two complementary categorization views. The \textbf{applications-focused} view describes target structures using familiar scientific terms (e.g., slab, interface, vacancy) and supports communication, dataset labeling, and discovery (Table~\ref{tab:categorization-science}). The \textbf{software-focused} view describes how those targets are realized in an implementation: as composable entities and operations that can be validated, reused, and assembled into complex structures (Tables~\ref{tab:categorization-software-core-entities}--\ref{tab:categorization-software-operations}).

        A compact M-CODE tag provides a short, stable identifier for frequent categories, supporting concise annotation of workflows and notebooks while keeping the detailed definition in schemas. The tag encodes a structural class (P for pristine, C for composite, D for defective, X for processed) together with dimensionality and a short category/variant suffix.

\section{Results}
\label{sec:results}

    This section summarizes the key artifacts produced by the M-CODE effort: the applications-focused categorization with compact tags, the software-focused view of reusable entities and operations, and the schema artifacts used for validation and reuse.
    
    \subsection{The Categorization and M-CODE tags}
    \label{subsec:results-categorization}
    
        The applications-focused view describes target structures using domain-science terminology while providing compact M-CODE tags (shown in table \ref{tab:categorization-science}) for concise annotation of workflows and datasets.
    
        \begin{table*}[!htbp]
            \label{tab:categorization-science}
            \centering
            \MateraTableSetup
            \begin{tabularx}{\textwidth}{|P{3.2cm}|P{1.0cm}|P{3.0cm}|Q|P{1.6cm}|}
                \hline
                \MateraTH{Domain} & \MateraTH{Dim.} & \MateraTH{Category} & \MateraTH{Variants} & \MateraTH{M-CODE} \\
                \hline
                Pristine Structures & 3D & Ideal Crystal & -- & \texttt{P-3D-CRY} \\
                & 2D & Monolayer & -- & \texttt{P-2D-MNL} \\
                & 2D & Slab & Simple & \texttt{P-2D-SLB-S} \\
                & 2D & Slab & Reconstructed & \texttt{P-2D-SLB-R} \\
                & 1D & Nanotape & -- & \texttt{P-1D-NTP} \\
                & 1D & Nanowire & -- & \texttt{P-1D-NWR} \\
                & 0D & Nanoparticle & -- & \texttt{P-0D-NPR} \\
                & 0D & Nanoribbon & -- & \texttt{P-0D-NRB} \\
                \hline
                Compound Pristine Structures & 2D & Heterostack & -- & \texttt{C-2D-HST} \\
                & 2D & Interface & Simple & \texttt{C-2D-INT-S} \\
                & 2D & Interface & ZSL & \texttt{C-2D-INT-Z} \\
                & 2D & Interface & Twisted Interface & \texttt{C-2D-INT-T} \\
                & 2D & Interface & Commensurate Lattice & \texttt{C-2D-INT-C} \\
                & 2D & Multi-Layer & -- & \texttt{C-2D-MLT} \\
                & 0D & Nanoribbons Interface & Rotation & \texttt{C-0D-INT-R} \\
                \hline
                Defective Structures & 3D & Amorphous & -- & \texttt{D-3D-AMO} \\
                & 2D & Adatom & -- & \texttt{D-2D-ADA} \\
                & 2D & Grain Boundary Planar & -- & \texttt{D-2D-GBP} \\
                & 2D & Island & -- & \texttt{D-2D-ISL} \\
                & 2D & Terrace & -- & \texttt{D-2D-TER} \\
                & 1D & Grain Boundary Linear & -- & \texttt{D-1D-GBL} \\
                & 0D & Defect Pair & -- & \texttt{D-0D-DFP} \\
                & 0D & Interstitial & -- & \texttt{D-0D-INT} \\
                & 0D & Substitution & -- & \texttt{D-0D-SUB} \\
                & 0D & Vacancy & -- & \texttt{D-0D-VAC} \\
                \hline
                Processed Structures & 3D & Perturbation & -- & \texttt{X-3D-PER} \\
                & 3D & Annealed Crystal & -- & \texttt{X-3D-ANL} \\
                & 2D & Passivated Surface & -- & \texttt{X-2D-PAS} \\
                & 1D & Passivated Edge & -- & \texttt{X-1D-PAS} \\
                & 0D & Passivated Edge & -- & \texttt{X-0D-PAS} \\
                & 0D & Slab Cutout & -- & \texttt{X-0D-CUT} \\
                \hline
            \end{tabularx}
            \caption{Applications-focused categorization of target structures with compact M-CODE tags. Columns: \textbf{Domain}-structural family (Pristine, Compound Pristine, Defective, Processed); \textbf{Dim.}-dimensionality (3D, 2D, 1D, 0D); \textbf{Category}-specific structure type; \textbf{Variants}-sub-types or specific configurations; \textbf{M-CODE}-compact tag encoding the categorization.}
        \end{table*}
    
        \begin{figure*}[!htbp]
            \label{fig:materials-grid}
            \centering
            \newcommand{\MateraGridCell}[3]{%
                \begin{minipage}[t]{0.23\textwidth}
                    \centering
                    \includegraphics[width=\linewidth,height=\linewidth]{#1}\\[2pt]
                    {\scriptsize\texttt{#2}}\\
                    {#3}
                \end{minipage}
            }
            {\setlength{\tabcolsep}{6pt}\renewcommand{\arraystretch}{1.2}
            \begin{tabular}{p{0.23\textwidth} p{0.23\textwidth} p{0.23\textwidth} p{0.23\textwidth}}
                \MateraGridCell{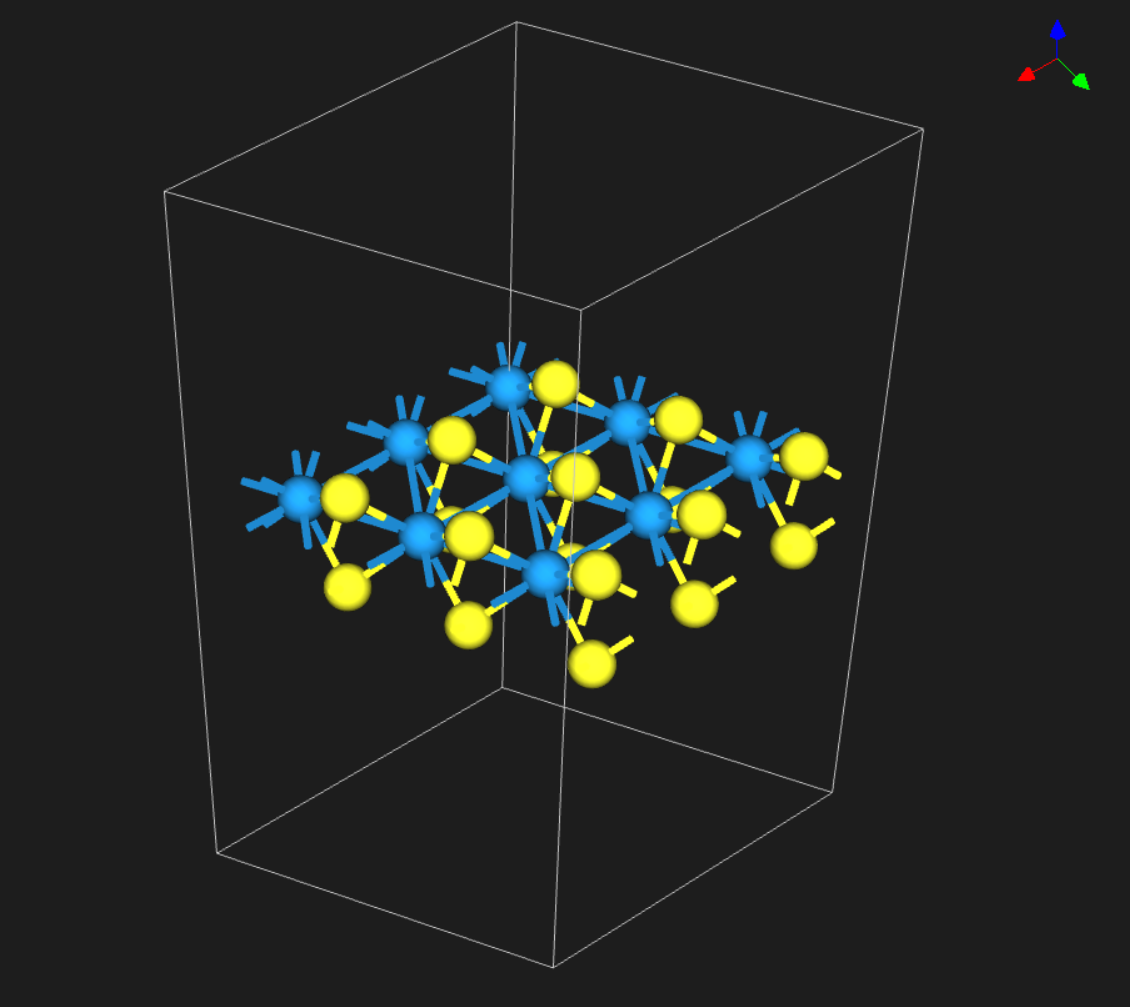}{P-2D-MNL}{Monolayer} &
                \MateraGridCell{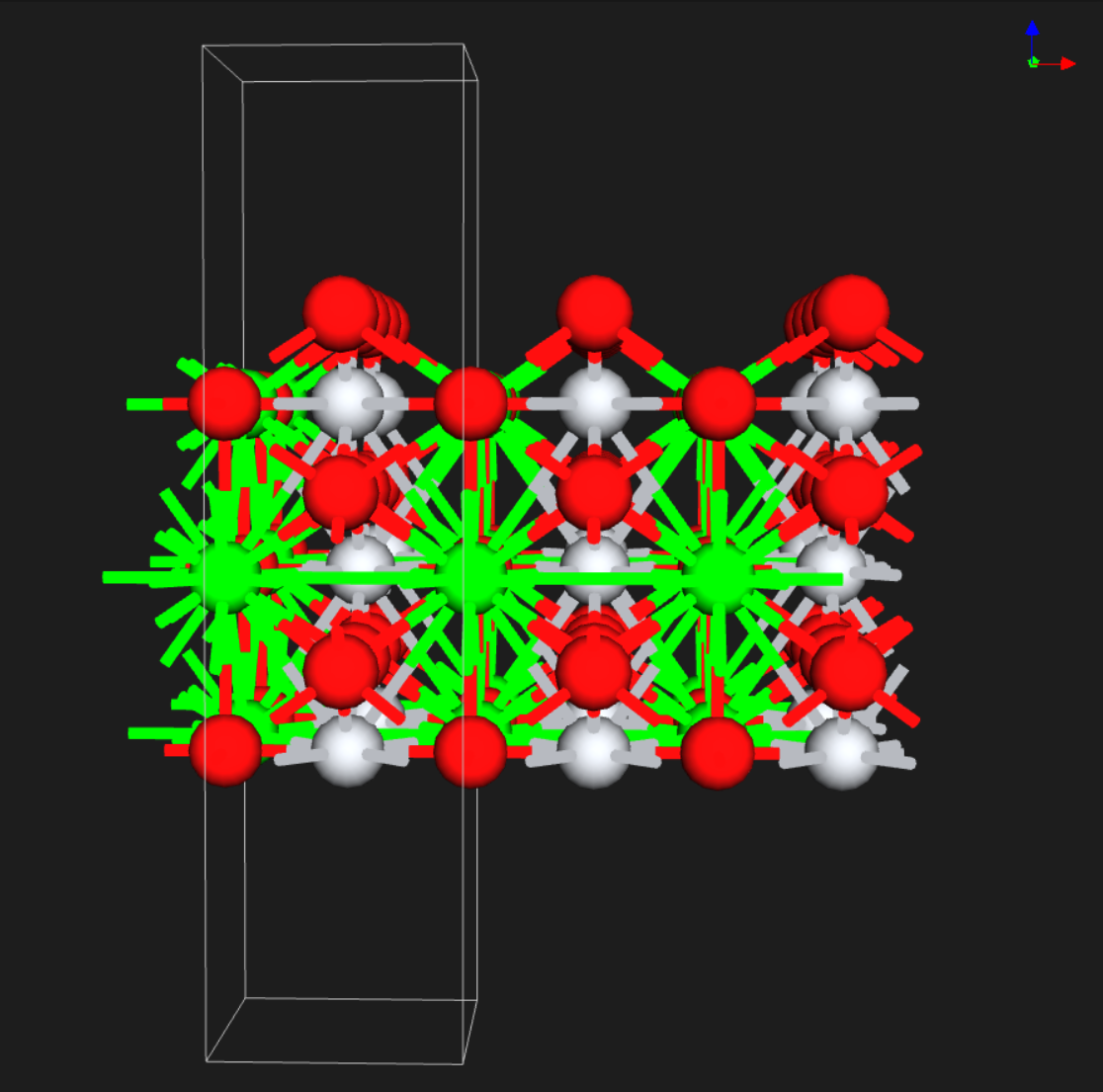}{P-2D-SLB-S}{Slab (simple)} &
                \MateraGridCell{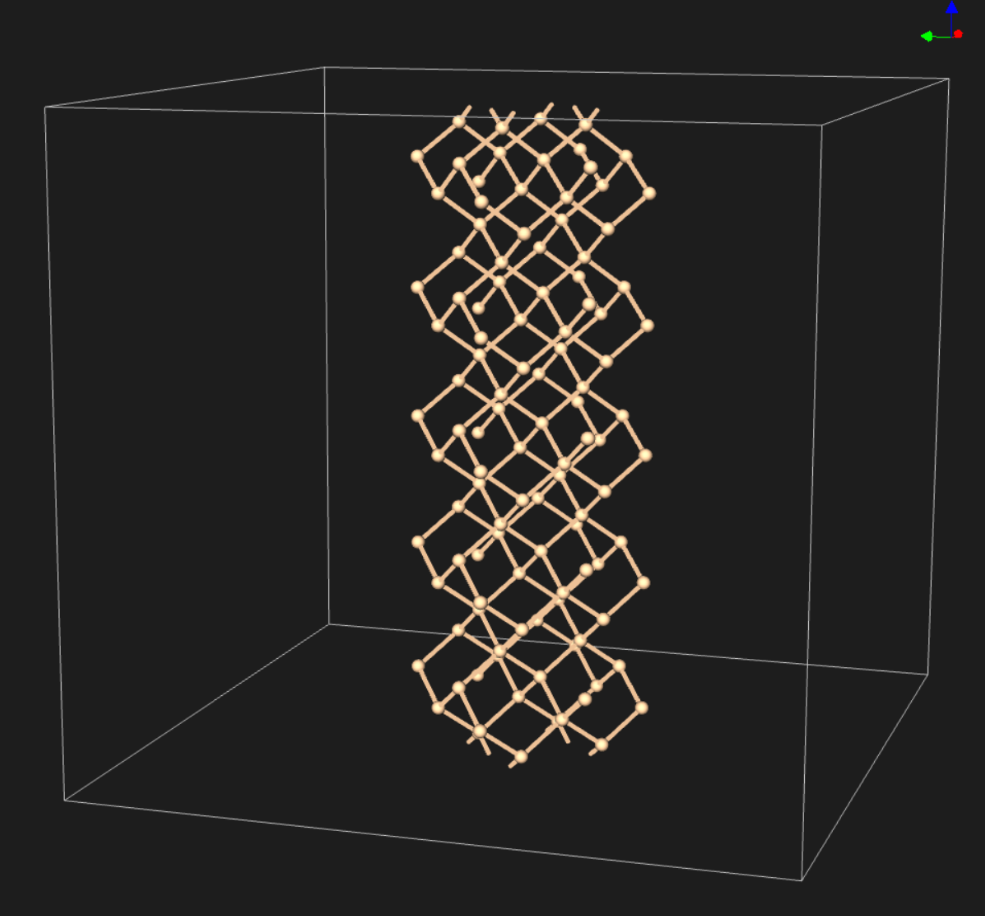}{P-1D-NWR}{Nanowire} &
                \MateraGridCell{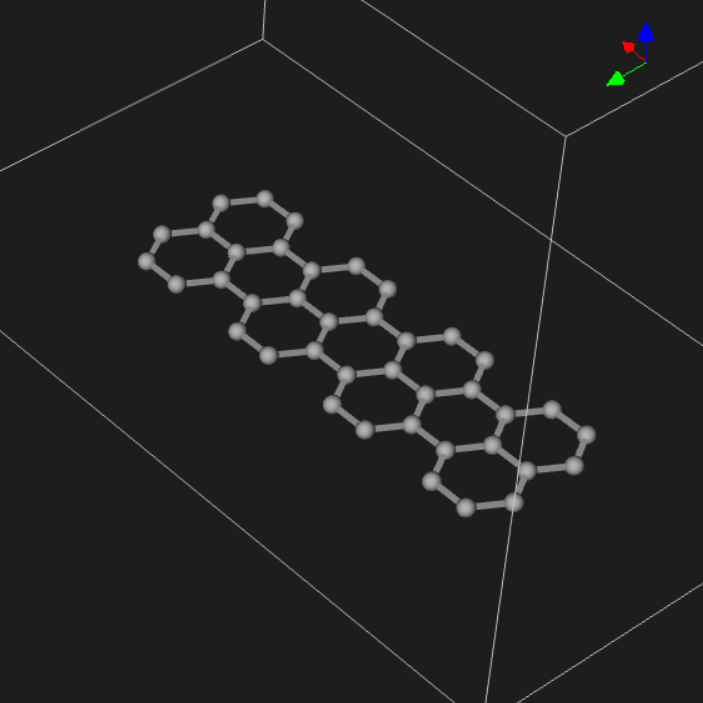}{P-0D-NRB}{Nanoribbon} \\
                \MateraGridCell{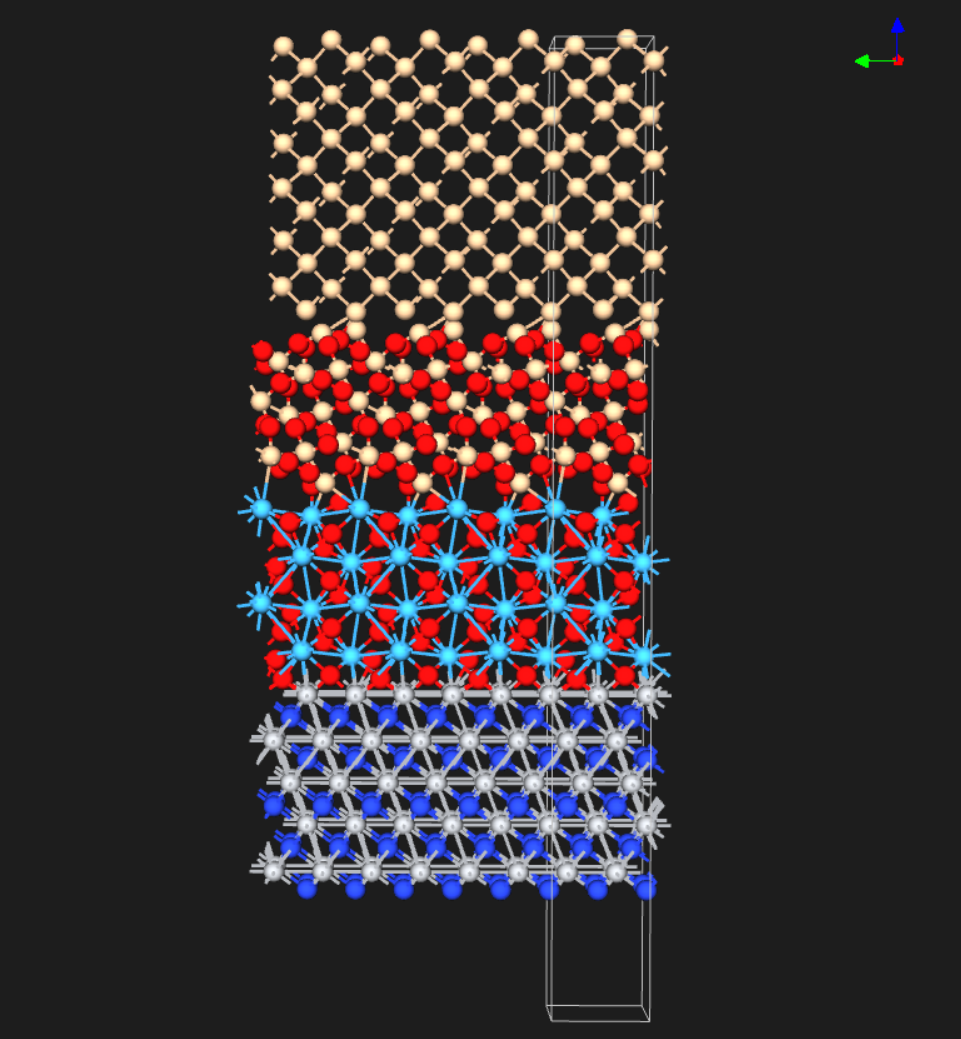}{C-2D-HST}{Heterostack} &
                \MateraGridCell{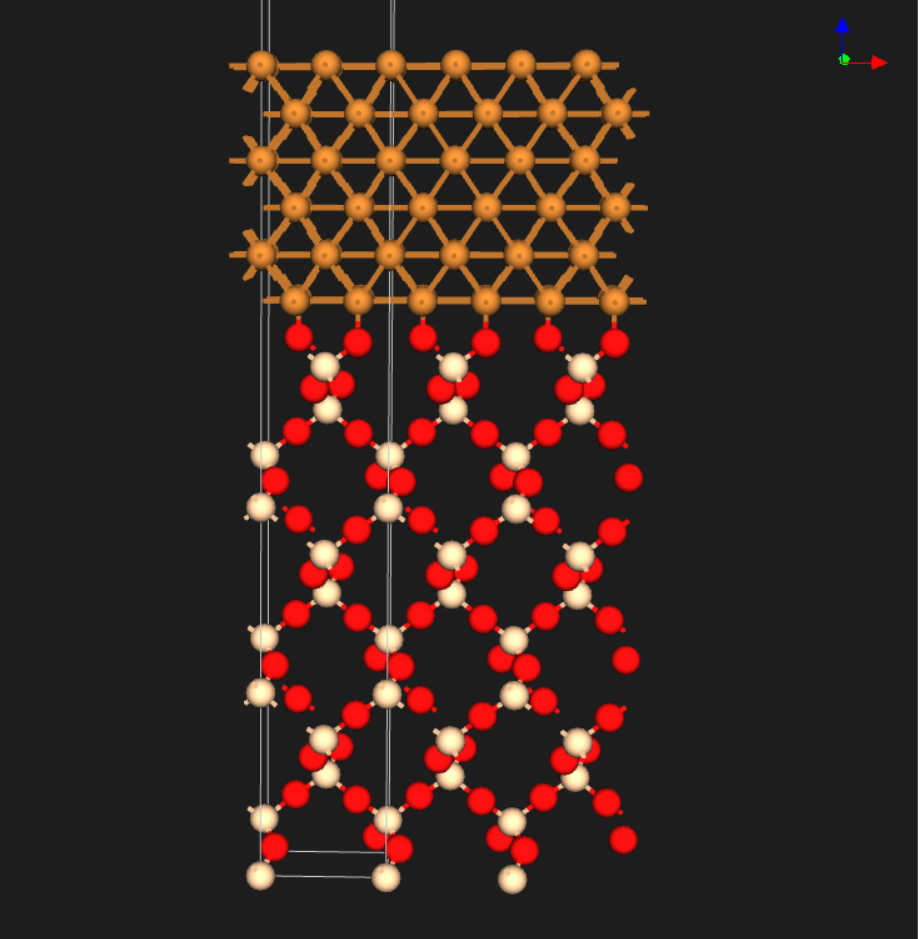}{C-2D-INT-Z}{Interface (ZSL)} &
                \MateraGridCell{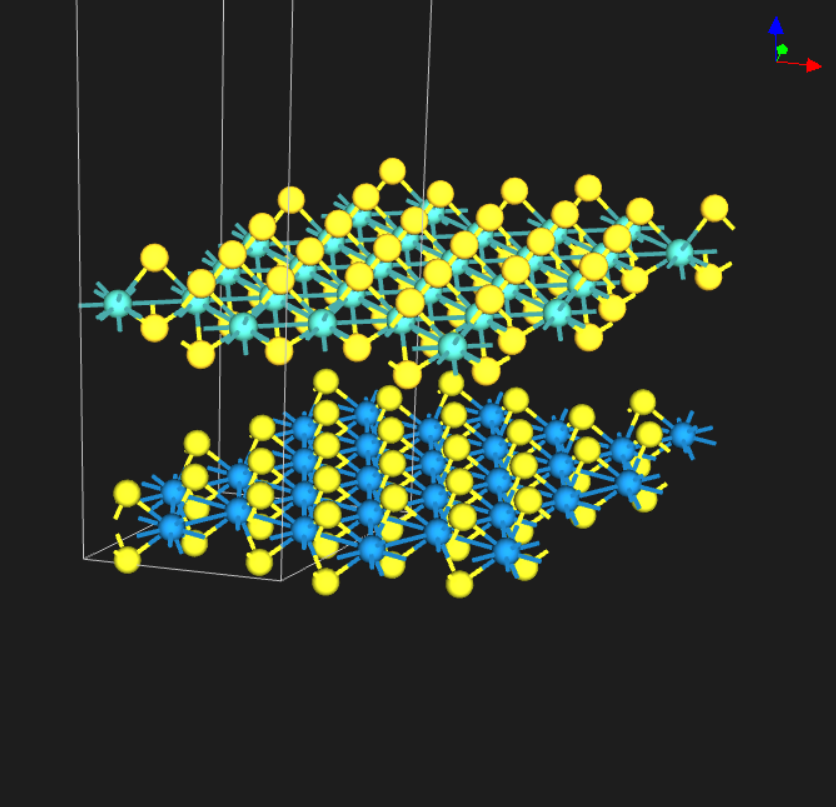}{C-2D-INT-T}{Interface (twisted/commensurate)} &
                \MateraGridCell{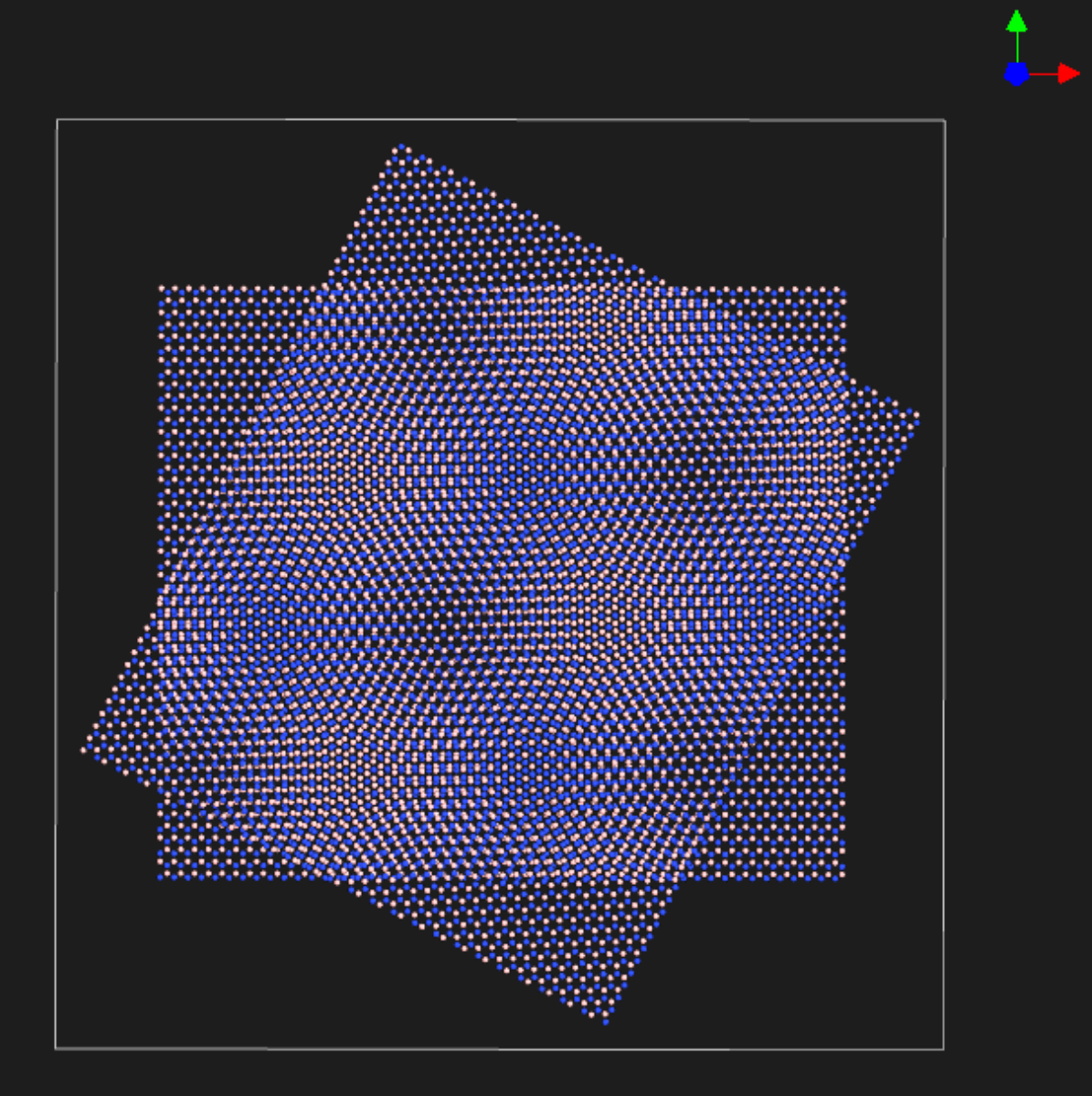}{C-0D-INT-R}{Nanoribbon interface} \\
                \MateraGridCell{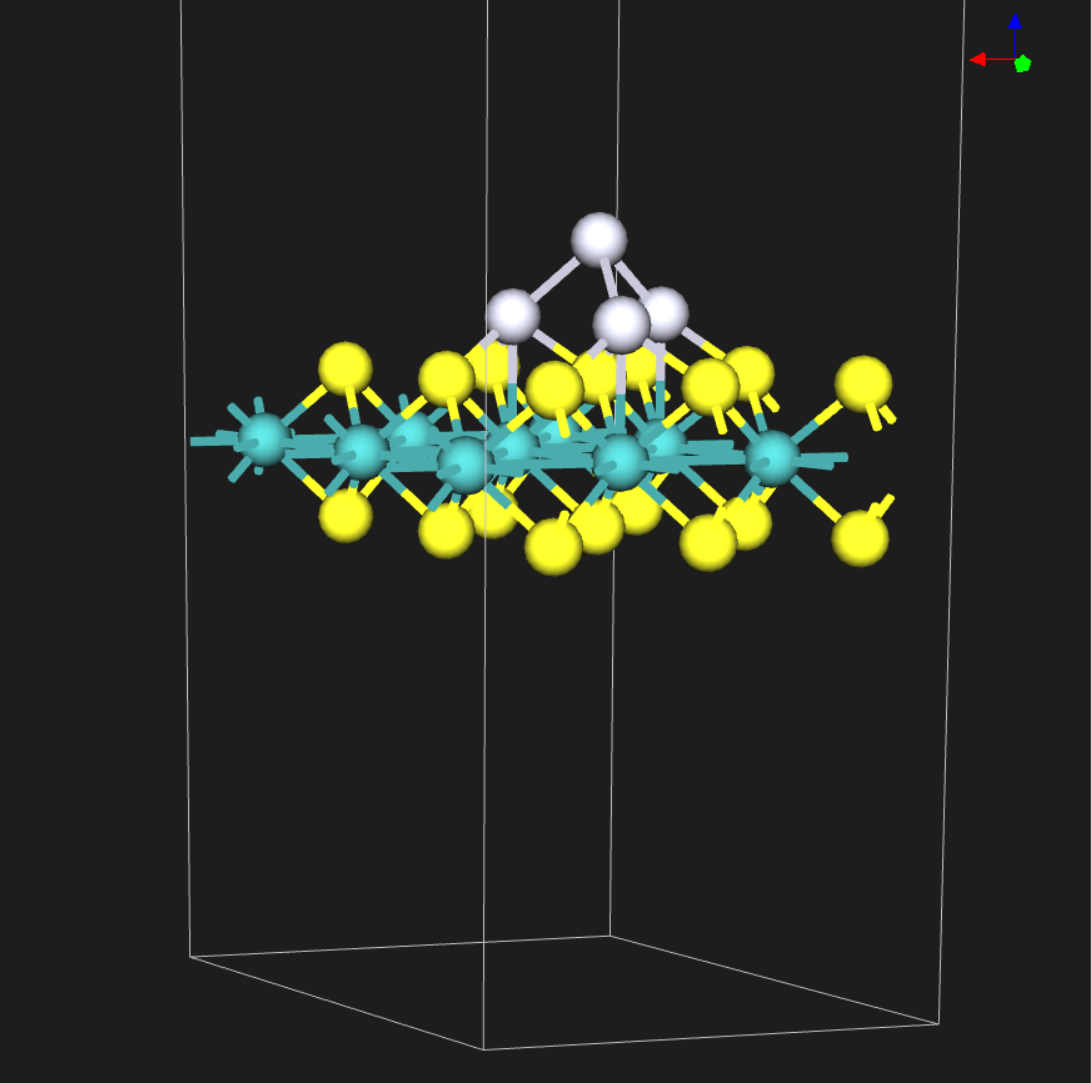}{D-2D-ADA}{Adatom} &
                \MateraGridCell{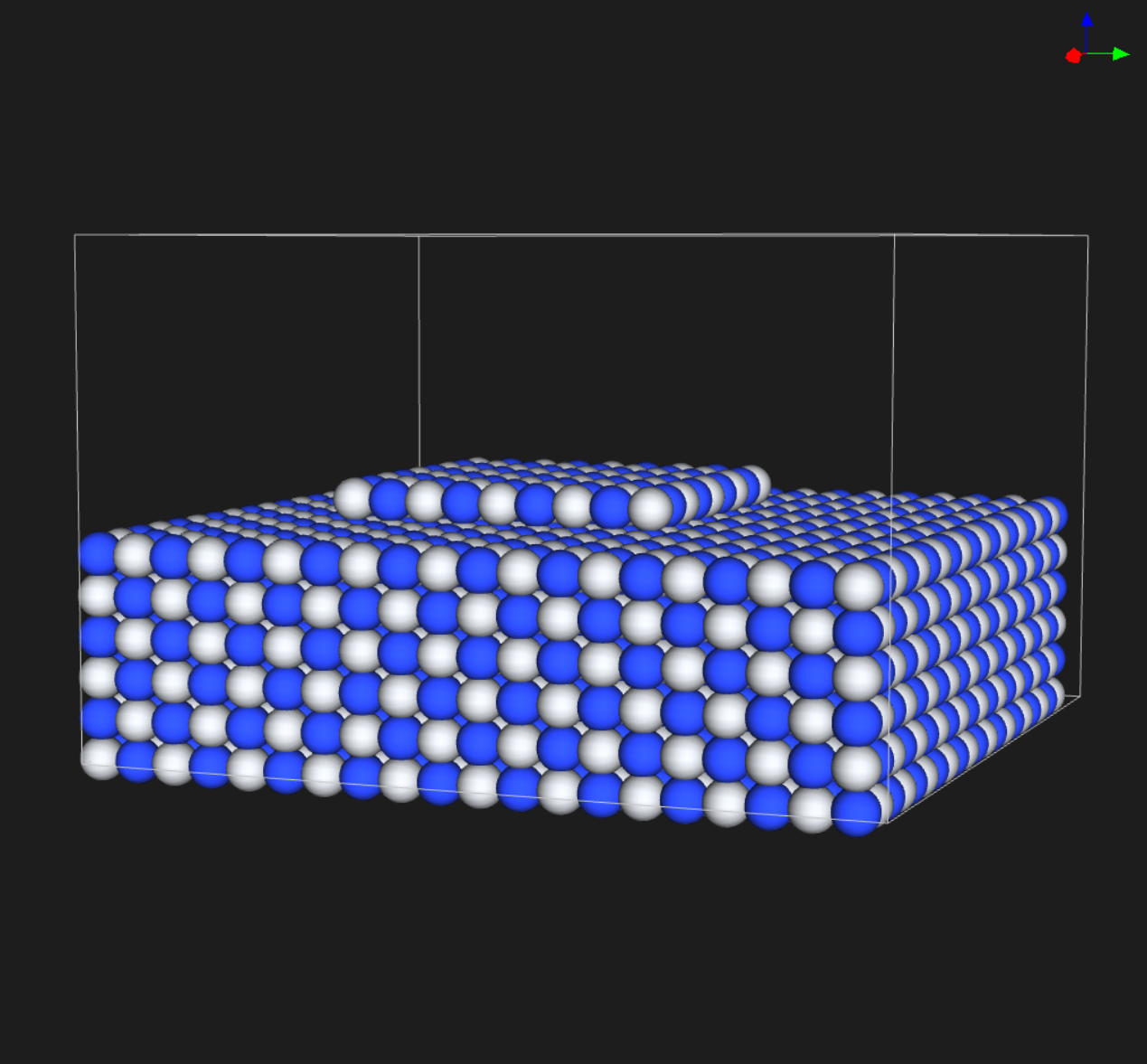}{D-2D-ISL}{Island} &
                \MateraGridCell{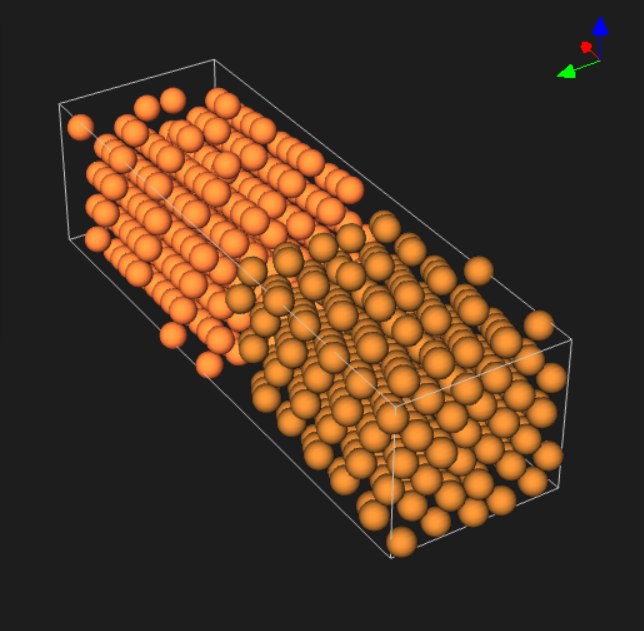}{D-2D-GBP}{Grain boundary} &
                \MateraGridCell{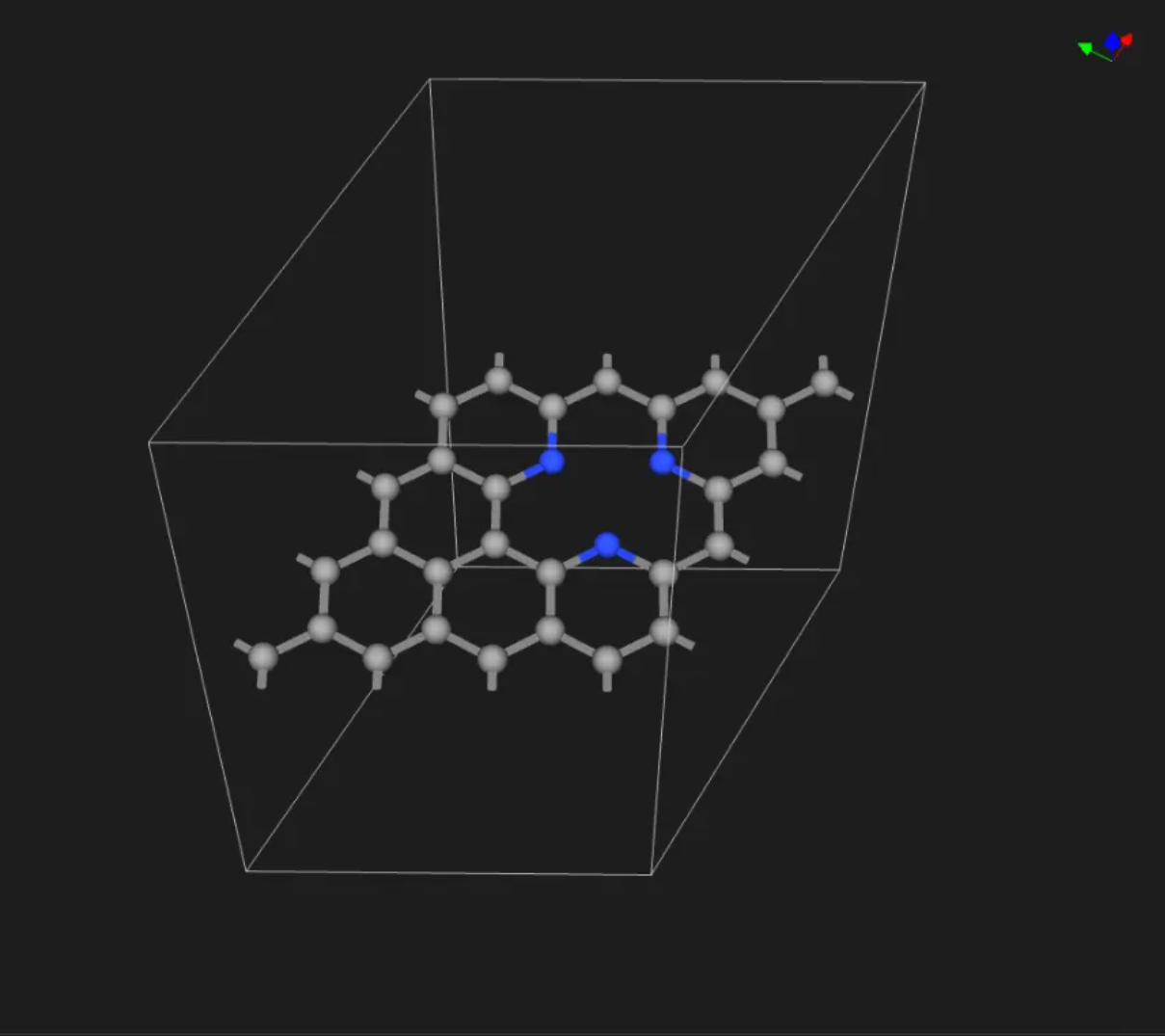}{D-0D-SUB}{Substitution} \\
                \MateraGridCell{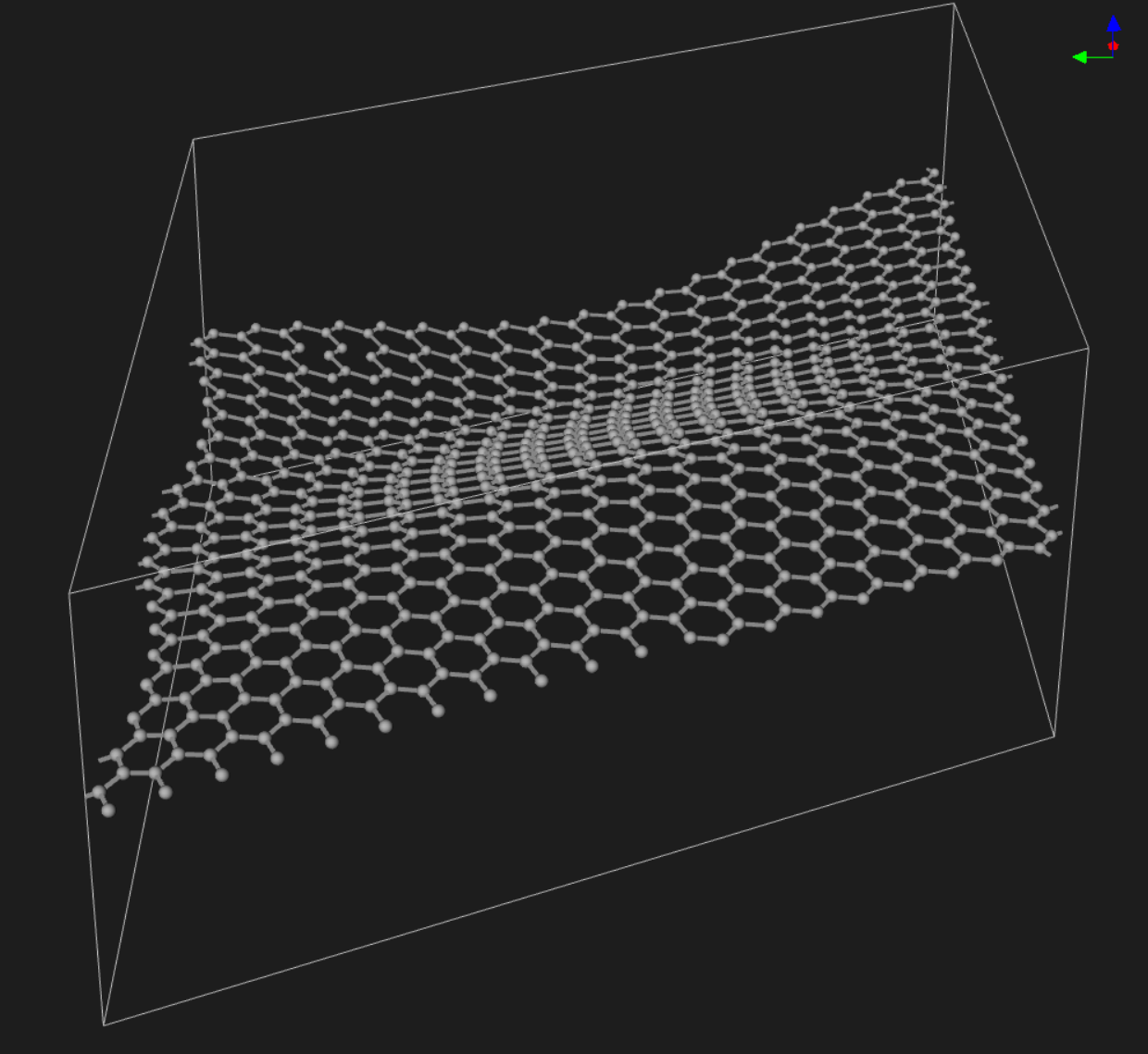}{X-3D-PER}{Perturbation} &
                \MateraGridCell{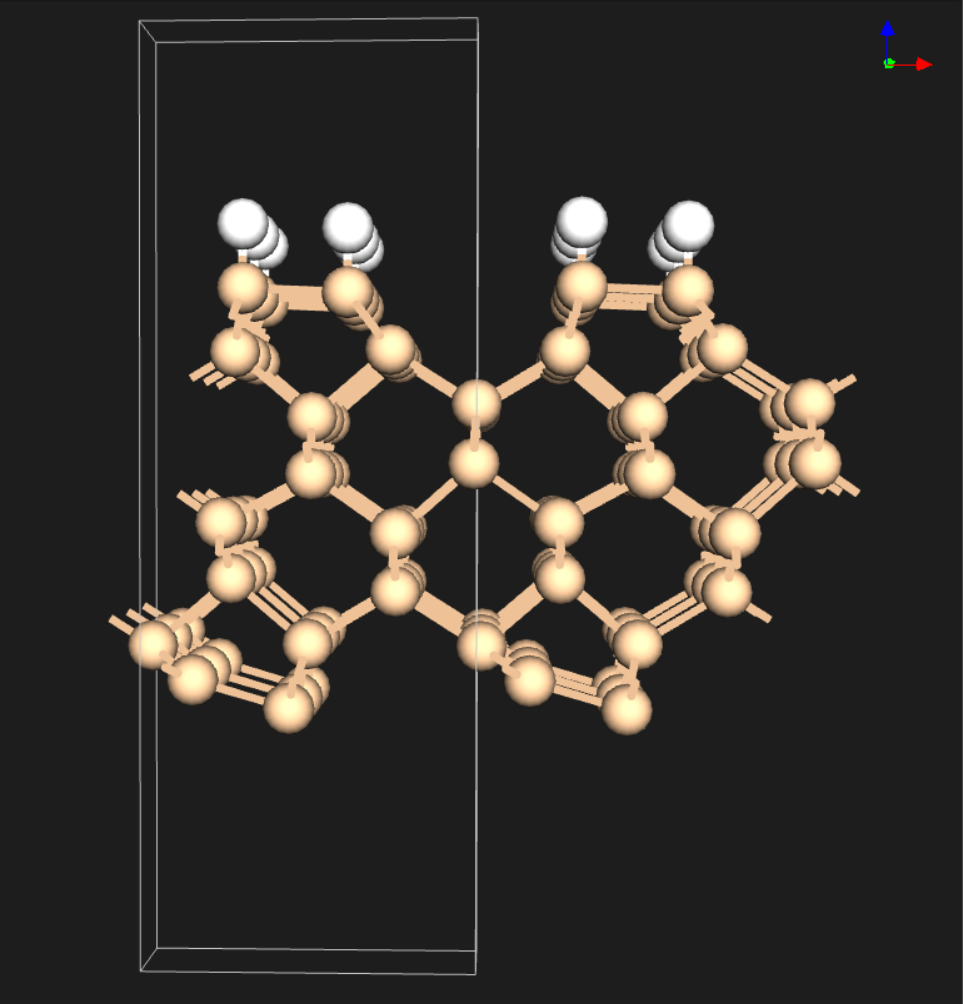}{X-2D-PAS}{Passivated surface} &
                \MateraGridCell{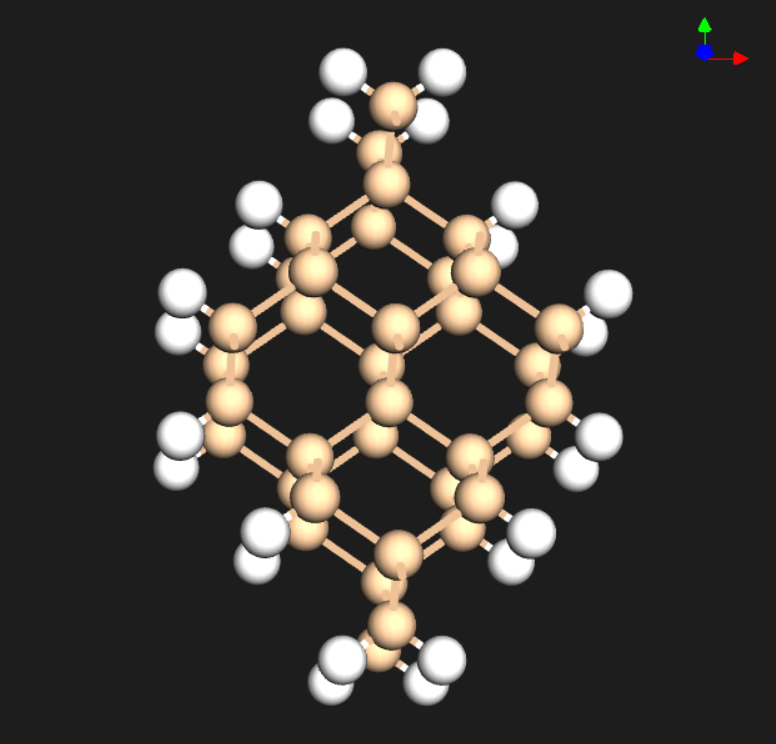}{X-1D-PAS}{Passivated edge} &
                \MateraGridCell{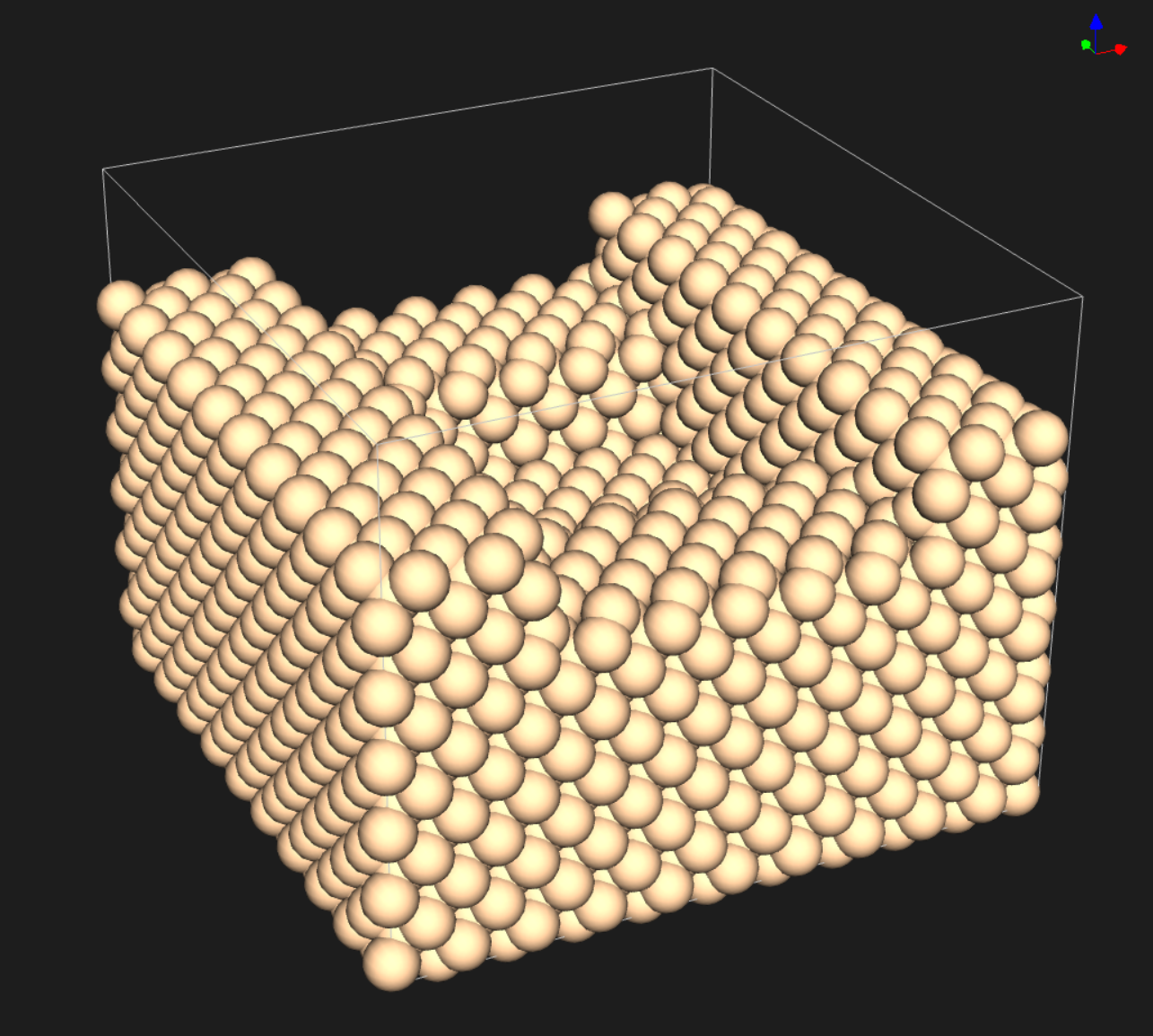}{X-0D-CUT}{Slab cutout} \\
            \end{tabular}}
            \caption{Representative target structure classes grouped by domain and annotated with the corresponding M-CODE tags. See Table \ref{tab:categorization-science}. \textbf{Pristine Structures} (top row): MoS$_2$ monolayer, SrTiO$_3$ slab, Si nanowire, and graphene nanoribbon. \textbf{Compound Pristine Structures} (second row): Si/SiO$_2$/HfO$_2$/TiN heterostack, Cu/SiO$_2$ interface (ZSL-matched), twisted MoS$_2$/WS$_2$ interface, and hBN/hBN nanoribbon interface. \textbf{Defective Structures} (third row): Pt adatom on WS$_2$, island on TiN(001), Cu(321)/($\bar{3}\bar{2}\bar{1}$) grain boundary, and nitrogen substitution in graphene. \textbf{Processed Structures} (bottom row): perturbed graphene, passivated Si surface, passivated Si nanowire edge, and Si box cutout.}
        \end{figure*}

    \clearpage

    \subsection{Schemas and examples}
    \label{subsec:results-schemas}
        M-CODE is implemented as JSON schemas, example entities, and reference manifests. The canonical artifacts are distributed as \texttt{mat3ra-esse}~\cite{mat3ra-esse-pypi-package} and can be used for validation, exchange, and provenance-aware regeneration of realistic structures. In ESSE, schemas are defined once and corresponding language bindings (e.g., Python and TypeScript interfaces) are generated automatically to keep implementations consistent. Stable versions are published as tagged releases and as package versions to support long-term reproducibility.

        \subsubsection{Schema and JSON example: interface}
        \label{subsubsec:results-interface}
            For readability, we show an abridged schema excerpt and a corresponding abridged JSON instance for a two-dimensional interface configuration. The canonical schemas are distributed with \texttt{mat3ra-esse}.
        
            Here's a schema for an interface, comprising a stack of two strained supercells and vacuum, with a specific xy\_shift:
\begin{lstlisting}[language=json]
{
  "$id": "materials-category/compound-pristine-structures/2d/interface/configuration",
  "title": "Interface Configuration Schema",
  "description": "A 2D interface between two slabs, optionally including vacuum, with a specified stacking direction.",
  "allOf": [{ "$ref": "operations/combinations/stack" }],
  "properties": {
    "stack_components": {
      "type": "array",
      "minItems": 2,
      "maxItems": 3,
      "items": [
        { "$ref": "pristine/2d/slab OR pristine/2d/slab_strained_supercell" },
        { "$ref": "pristine/2d/slab OR pristine/2d/slab_strained_supercell" },
        { "$ref": "core/2d/vacuum" }
      ]
    },
    "direction": { "default": "z" },
    "xy_shift": { "default": [0.0, 0.0] }
  }
}

\end{lstlisting}
        
            And here's the corresponding example for a Ni-Graphene interface separated by 10 \r{A}.
\begin{lstlisting}[language=json]
{
  "stack_components": [
    { "type": "slab", "material": "Graphene", "miller_indices": [0, 0, 1], "..." : "..." },
    { "type": "slab", "material": "Ni",       "miller_indices": [0, 0, 1], "..." : "..." },
    { "type": "vacuum", "direction": "z", "size": 10.0 }
  ],
  "direction": "z",
  "xy_shift": [0.0, 0.0]
}

\end{lstlisting}
        
        \subsubsection{Schema and JSON example: substitutional point defect}
        \label{subsubsec:results-substitution}
            The substitutional point defect schema below embeds a point-defect base configuration that uses a merge operation, and specializes it by constraining the merged components.
        
            Here's a schema for a point defect where a 3d crystal/bulk material is merged with a specific defect site with "REPLACE" as the type of the merge method:
\begin{lstlisting}[language=json]
{
  "$id": "materials-category/defective-structures/0d/point-defect/substitutional",
  "title": "Substitutional Point Defect Schema",
  "description": "A substitutional point defect configuration where an atom at a crystal site is replaced with a different element.",
  "allOf": [{
    "title": "Point Defect Base Configuration Schema",
    "allOf": [{ "$ref": "operations/combinations/merge" }],
    "properties": {
      "merge_method": { "default": "REPLACE" }
    },
    "required": ["merge_components", "merge_method"]
  }],
  "properties": {
    "merge_components": {
      "items": [
        { "$ref": "core/3d/crystal" },
        { "$ref": "aux/0d/point_defect_site" }
      ]
    }
  }
}

\end{lstlisting}
        
            And here's the resulting nitrogen substitution defect in graphene:
\begin{lstlisting}[language=json]
{
  "merge_components": [
    { "type": "crystal", "material": "Graphene", "..." : "..." },
    { "type": "point_defect_site", "coordinate": [0.0, 0.0, 0.0], "chemical_element": "N" }
  ],
  "merge_method": "REPLACE"
}

\end{lstlisting}
        
        \subsubsection{Schema and JSON example: nanoribbon}
        \label{subsubsec:results-nanoribbon}
            The nanoribbon schema below embeds a nanotape schema: a nanoribbon is defined as a nanotape stacked with vacuum in the $x$ direction, while the nanotape itself is defined as lattice lines stacked with vacuum in the $y$ direction.
        
            Here's a nanoribbon configuration where we first stack crystal lattice lines with vacuum in one direction (Y), then the result is stacked with vacuum in the other direction (X):
\begin{lstlisting}[language=json]
{
  "$id": "materials-category/pristine-structures/2d/nanoribbon",
  "title": "Nanoribbon Configuration Schema",
  "description": "A nanoribbon unit cell consisting of nanotape and vacuum stacked in X direction.",
  "allOf": [{ "$ref": "operations/combinations/stack" }],
  "properties": {
    "stack_components": [
      {
        "title": "NanoTape Configuration Schema",
        "allOf": [{ "$ref": "operations/combinations/stack" }],
        "properties": {
          "stack_components": [
            { "$ref": "reusable/1d/crystal_lattice_lines_unique_repeated" },
            { "$ref": "core/2d/vacuum" }
          ],
          "direction": { "default": "y" }
        }
      },
      { "$ref": "core/2d/vacuum" }
    ],
    "direction": { "default": "x" }
  }
}

\end{lstlisting}
        
            And here's the resulting nanoribbon with 10 \r{A} aligned along Z of vacuum on each side:
\begin{lstlisting}[language=json]
{
  "stack_components": [
    {
      "stack_components": [
        { "type": "crystal_lattice_lines_unique_repeated", "..." : "..." },
        { "type": "vacuum", "direction": "y", "size": 10.0 }
      ],
      "direction": "y"
    },
    { "type": "vacuum", "direction": "x", "size": 10.0 }
  ],
  "direction": "x"
}

\end{lstlisting}

    \subsection{Provenance}
    \label{subsec:results-provenance}
    
        This composition model enables systematic construction while keeping provenance explicit and machine-readable, stored as metadata in created materials. To facilitate the operation of software systems that use the convention, it is important to allow the storage of additional information (e.g., timestamps, environment metadata, and logs), so that schemas are permissive of the addition of extra keys.

        For example, a generated slab can carry build provenance as permissive JSON metadata, describing the configuration components and build parameters:
        \clearpage
\begin{lstlisting}[language=json]
{
    "metadata": {
        "build": [
            {
                "configuration": {
                    "stack_components": [
                        { "$ref": "pristine/2d/slab OR pristine/2d/slab_strained_supercell" },
                        { "$ref": "core/2d/vacuum" }
                    ],
                    "direction": "z",
                    "gaps": [],
                    "type": "SlabConfiguration"
                },
                "build_parameters": {
                    "use_orthogonal_c": true,
                    "xy_supercell_matrix": [
                        [1, 0],
                        [0, 1]
                    ]
                }
            }
        ]
    }
}
\end{lstlisting}

\section{Discussion}
\label{sec:discussion}

    M-CODE complements existing standards and infrastructure that address adjacent layers of the materials data ecosystem. OPTIMADE standardizes a common API and response conventions for querying and serving materials entries across databases \cite{optimade}. NOMAD focuses on archival, normalization, and dissemination of computational materials data and metadata \cite{nomad}. AiiDA provides an execution framework with workflow-level provenance tracking \cite{pizzi2016aiida}. In contrast, M-CODE targets the intermediate structure-definition layer: a shared, machine-readable vocabulary for structure classes, build inputs, and provenance-aware construction steps that can be embedded in datasets and mapped consistently to implementations across languages.

    M-CODE is motivated by a practical constraint in data-driven materials modeling: the complexity of materials science is in its extreme diversity. Useful structures are rarely ideal bulk crystals, containing a myriad of intrinsic details. While many datasets, benchmarks, and codebases assume ideal structures, with M-CODE, our attempt is to avoid such over-simplification and be able to apply the ``divide-and-conquer'' mentality to manage the diversity of different use cases. A shared categorization helps specify what structures are included, how they relate, and which transformations and parameters produced them.

    \subsection{Object-oriented design and reusability}
    \label{subsec:discussion-oop}
        A central design choice in M-CODE is the use of object-oriented principles: entities, operations, and configurations are defined as composable, self-contained units. This modularity has practical consequences beyond the categorization itself. The same entity and operation definitions that describe a slab or an interface in a JSON schema can be mapped directly to classes and methods in a software codebase, to components in a user interface, to features in an AI/ML model, and to folder and file naming conventions in a data repository. When a new structure class is added to the categorization, the corresponding software class, UI component, and data directory follow from the same definition, reducing duplication and keeping all digital artifacts consistent. This reusability lowers the cost of maintaining and extending the framework as the number of supported structure types grows.

    \subsection{Category-aware modeling and characterization}
    \label{subsec:discussion-category-aware}
        Because each structure carries its category and provenance, downstream tools can select appropriate modeling or characterization approaches automatically. For example, a slab tagged \texttt{P-2D-SLB-S} can be routed to a surface-energy workflow with dipole corrections enabled, while an interface tagged \texttt{C-2D-INT-Z} can trigger a lattice-matching pre-processing step followed by an adhesion-energy calculation. Similarly, a defective structure tagged \texttt{D-0D-VAC} can be directed to a defect-formation-energy workflow that requires a pristine reference. This category-to-workflow mapping reduces manual setup, prevents mismatches between structure type and simulation parameters, and makes it straightforward to apply the same protocol consistently across large datasets.

    \subsection{FAIR principles and AI-readiness}
    \label{subsec:discussion-fair}
        The ontology underlying M-CODE directly supports the FAIR data principles \cite{wilkinson2016fair}: structures are \textbf{Findable} through compact M-CODE tags and category metadata; \textbf{Accessible} via open JSON schemas distributed with the reference implementation; \textbf{Interoperable} because the same entity and operation vocabulary is shared across tools, languages, and databases; and \textbf{Reusable} because provenance metadata records the full sequence of inputs and transformations needed to regenerate any structure. For AI/ML applications, this means that training sets can be filtered, balanced, and versioned by structural category, and that model performance can be evaluated per category rather than only in aggregate. The explicit provenance also makes it possible to detect and correct data leakage between training and test splits when structures share a common parent material or transformation path.

    \subsection{Implications for data and AI/ML}
    \label{subsec:discussion-ai-ml}
        Dataset quality is shaped not only by the underlying electronic-structure method but also by how structures are defined, transformed, and curated. Many benchmarks emphasize structures that are easy to enumerate and standardize, which can lead to models that perform well in-distribution but fail to transfer to heterogeneous, defect-rich, and interface-dominated settings \cite{ward2016ml-framework, isayev2017ml-descriptors}. A compact, provenance-aware categorization helps specify distributions of structures explicitly and supports regeneration of training and test sets as assumptions evolve, complementing broader efforts on reproducibility and traceability \cite{DFTReproducibility2016lejaeghere, EB-ARX-2023}.

    \subsection{Practical considerations and limitations}
    \label{subsec:discussion-limitations}
        A categorization is necessarily an evolving artifact. New structure classes, variants, and parameterizations emerge as modeling targets shift, and different communities may prefer different abstractions. The current version of M-CODE reflects the structure types most commonly encountered in our workflows and may not yet cover all domains equally (e.g., solid electrolyte interphases, amorphous alloys, or multi-scale models that span atomistic and continuum descriptions). The methodology is most effective when paired with clear dataset design goals and transparent reporting of generation parameters so that disagreements about definitions can be surfaced, discussed, and resolved in a reproducible way.

    \subsection{Open-source development and community input}
    \label{subsec:discussion-community}
        M-CODE is released as open-source schemas and examples and is informed by our experience building materials generation tooling for scientific workflows since 2015 \cite{EB-ARX-02-2019}. The open-source approach is essential: a categorization standard is only as useful as the community that adopts and extends it. By publishing schemas, example instances, and reference code under a permissive license, we lower the barrier for others to validate the definitions against their own use cases, propose additions for missing structure classes, and integrate M-CODE tags into their own databases and pipelines. We expect that community feedback will refine the hierarchy, expand coverage of structure classes and variants, and improve interoperability with other standards and databases.

    \subsection{Versioning and governance}
    \label{subsec:discussion-versioning}

        Because tags and schemas are intended for long-term reuse, changes must be explicit and traceable. We version schema artifacts and tag definitions, publish release notes describing additions and breaking changes, and use deprecation periods to transition away from renamed or redefined classes. This supports stable dataset labels, reproducible regeneration of structures, and transparent evolution of the categorization as new use cases emerge.
\section{Conclusion}
\label{sec:conclusion}

    This work introduced M-CODE as a compact categorization system for realistic low-dimensional and heterogeneous structures used in data-driven materials modeling. We described how M-CODE links domain-science terminology to reusable entities and operations and how it supports provenance-aware regeneration through schema-validated configurations and transformations. By making structure classes and variants explicit and standardized, M-CODE aims to improve dataset design, benchmarking, and interoperability across tools and communities.

\section*{Acknowledgements}
This work was supported in part by NIST 70NANB24H205.

During the preparation of this work, the author(s) used large language models (LLMs) to assist with drafting and editing portions of this manuscript. After using this tool/service, the author(s) reviewed and edited the content as needed and take(s) full responsibility for the content of the published article.

\section*{Data and Software availability}
M-CODE and the associated schemas are maintained as open-source data standards. The reference implementation is accessible as a package on GitHub and distributed via PyPI as \texttt{mat3ra-esse}. Related ecosystem packages and notebooks are available on GitHub and PyPI, including \texttt{mat3ra-standata}, \texttt{mat3ra-made}, and \texttt{mat3ra-api-examples} \cite{mat3ra-esse-pypi-package, mat3ra-standata-pypi-package, mat3ra-made-pypi-package, mat3ra-api-examples-pypi-package}.

\bibliographystyle{unsrtnat}

\bibliography{references}

@manual{exabyteESSEGithubRepo,
    URL= {https://github.com/exabyte-io/esse},
    title = {mat3ra-esse: schemas and examples},
    year = 2026
}

@article{EB-ARX-02-2019,
  author    = {Bazhirov, T.},
  title     = {Data-centric online ecosystem for digital materials science},
  journal   = {arXiv preprint arXiv:1902.10838},
  year      = {2019},
  url       = {https://arxiv.org/abs/1902.10838}
}

@article{EB-ARX-2023,
  author    = {Dean, J. and Scheffler, M. and Purcell, T. A. R. and Barabash, S. V. and Bhowmik, R. and Bazhirov, T.},
  title     = {Interpretable machine learning for materials design},
  journal   = {Journal of Materials Research},
  volume    = {38},
  pages     = {4477--4496},
  year      = {2023},
  doi       = {10.1557/s43578-023-01164-w},
  url       = {https://doi.org/10.1557/s43578-023-01164-w}
}

@manual{mat3ra-com-website,
    organization = {Mat3ra.com},
    title = {Mat3ra.com},
    url= {https://www.mat3ra.com/},
    year = {2026},
}

@manual{mat3ra-esse-pypi-package,
    organization = {Mat3ra.com},
    title = {mat3ra-esse},
    url= {https://pypi.org/project/mat3ra-esse/},
    year = {2026},
}

@manual{mat3ra-esse-npm-package,
    organization = {Mat3ra.com},
    title = {@mat3ra/esse},
    url= {https://www.npmjs.com/package/@mat3ra/esse},
    year = {2026},
}

@manual{mat3ra-standata-pypi-package,
    organization = {Mat3ra.com},
    title = {mat3ra-standata},
    url= {https://pypi.org/project/mat3ra-standata/},
    year = {2026},
}

@manual{mat3ra-made-pypi-package,
    organization = {Mat3ra.com},
    title = {mat3ra-made},
    url= {https://pypi.org/project/mat3ra-made/},
    year = {2026},
}

@manual{mat3ra-api-examples-pypi-package,
    organization = {Mat3ra.com},
    title = {mat3ra-api-examples},
    url= {https://pypi.org/project/mat3ra-api-examples/},
    year = {2026},
}

@article{choudhary2020joint,
  title={The joint automated repository for various integrated simulations (JARVIS) for data-driven materials design},
  author={Choudhary, Kamal and Garrity, Kevin F and Reid, Andrew CE and DeCost, Brian and Biacchi, Adam J and Hight Walker, Angela R and Trautt, Zachary and Hattrick-Simpers, Jason and Kusne, A Gilad and Centrone, Andrea and others},
  journal={npj computational materials},
  volume={6},
  number={1},
  pages={173},
  year={2020},
  publisher={Nature Publishing Group UK London}
}

@article {DFTReproducibility2016lejaeghere,
	author = {Lejaeghere, Kurt and Bihlmayer, Gustav and Bj{\"o}rkman, Torbj{\"o}rn and Blaha, Peter and Bl{\"u}gel, Stefan and Blum, Volker and Caliste, Damien and Castelli, Ivano E. and Clark, Stewart J. and Dal Corso, Andrea and de Gironcoli, Stefano and Deutsch, Thierry and Dewhurst, John Kay and Di Marco, Igor and Draxl, Claudia and Du{\l}ak, Marcin and Eriksson, Olle and Flores-Livas, Jos{\'e} A. and Garrity, Kevin F. and Genovese, Luigi and Giannozzi, Paolo and Giantomassi, Matteo and Goedecker, Stefan and Gonze, Xavier and Gr{\r a}n{\"a}s, Oscar and Gross, E. K. U. and Gulans, Andris and Gygi, Fran{\c c}ois and Hamann, D. R. and Hasnip, Phil J. and Holzwarth, N. A. W. and Iu{\c s}an, Diana and Jochym, Dominik B. and Jollet, Fran{\c c}ois and Jones, Daniel and Kresse, Georg and Koepernik, Klaus and K{\"u}{\c c}{\"u}kbenli, Emine and Kvashnin, Yaroslav O. and Locht, Inka L. M. and Lubeck, Sven and Marsman, Martijn and Marzari, Nicola and Nitzsche, Ulrike and Nordstr{\"o}m, Lars and Ozaki, Taisuke and Paulatto, Lorenzo and Pickard, Chris J. and Poelmans, Ward and Probert, Matt I. J. and Refson, Keith and Richter, Manuel and Rignanese, Gian-Marco and Saha, Santanu and Scheffler, Matthias and Schlipf, Martin and Schwarz, Karlheinz and Sharma, Sangeeta and Tavazza, Francesca and Thunstr{\"o}m, Patrik and Tkatchenko, Alexandre and Torrent, Marc and Vanderbilt, David and van Setten, Michiel J. and Van Speybroeck, Veronique and Wills, John M. and Yates, Jonathan R. and Zhang, Guo-Xu and Cottenier, Stefaan},
	title = {Reproducibility in density functional theory calculations of solids},
	volume = {351},
	number = {6280},
	year = {2016},
	doi = {10.1126/science.aad3000},
	publisher = {American Association for the Advancement of Science},
	issn = {0036-8075},
	URL = {http://science.sciencemag.org/content/351/6280/aad3000},
	journal = {Science}
}

@article{larsen2017atomic,
  title={The atomic simulation environment—a Python library for working with atoms},
  author={Larsen, Ask Hjorth and Mortensen, Jens J{\o}rgen and Blomqvist, Jakob and Castelli, Ivano E and Christensen, Rune and Du{\l}ak, Marcin and Friis, Jesper and Groves, Michael N and Hammer, Bj{\o}rk and Hargus, Cory and others},
  journal={Journal of Physics: Condensed Matter},
  volume={29},
  number={27},
  pages={273002},
  year={2017},
  publisher={IOP Publishing}
}

@article{jain2013materialsproject,
    title={Commentary: The Materials Project: A materials genome approach to accelerating materials innovation},
    author={Jain, Anubhav and Ong, Shyue Ping and Hautier, Geoffroy and Chen, Wei and Richards, William Davidson and Dacek, Stephen and Cholia, Shreyas and Gunter, Dan and Skinner, David and Ceder, Gerbrand and others},
    journal={Apl Materials},
    volume={1},
    number={1},
    pages={011002},
    year={2013},
    publisher={AIP}
}

@article{ong2013python,
  title={Python Materials Genomics (pymatgen): A robust, open-source python library for materials analysis},
  author={Ong, Shyue Ping and Richards, William Davidson and Jain, Anubhav and Hautier, Geoffroy and Kocher, Michael and Cholia, Shreyas and Gunter, Dan and Chevrier, Vincent L and Persson, Kristin A and Ceder, Gerbrand},
  journal={Computational Materials Science},
  volume={68},
  pages={314--319},
  year={2013},
  publisher={Elsevier}
}

@article{curtarolo2012aflowlib,
  title={AFLOWLIB. ORG: A distributed materials properties repository from high-throughput ab initio calculations},
  author={Curtarolo, Stefano and Setyawan, Wahyu and Wang, Shidong and Xue, Junkai and Yang, Kesong and Taylor, Richard H and Nelson, Lance J and Hart, Gus LW and Sanvito, Stefano and Buongiorno-Nardelli, Marco and others},
  journal={Computational Materials Science},
  volume={58},
  pages={227--235},
  year={2012},
  publisher={Elsevier}
}

@article{pizzi2016aiida,
  title={AiiDA: automated interactive infrastructure and database for computational science},
  author={Pizzi, Giovanni and Cepellotti, Andrea and Sabatini, Riccardo and Marzari, Nicola and Kozinsky, Boris},
  journal={Computational Materials Science},
  volume={111},
  pages={218--230},
  year={2016},
  publisher={Elsevier}
}

@article{saal2013openQMD,
  title={Materials design and discovery with high-throughput density functional theory: the open quantum materials database (OQMD)},
  author={Saal, James E and Kirklin, Scott and Aykol, Muratahan and Meredig, Bryce and Wolverton, Christopher},
  journal={Jom},
  volume={65},
  number={11},
  pages={1501--1509},
  year={2013},
  publisher={Springer}
}

@manual {nomad,
    organization = {NOMAD Laboratory},
    url= {https://www.nomad-coe.eu/},
    title = {NOMAD laboratory: A European Centre of Excellence},
    year = 2018
}

@manual{optimade,
    organization = {OPTIMADE Consortium},
    title = {OPTIMADE: Open Databases Integration for Materials Design},
    url= {https://www.optimade.org/},
    year = 2026
}

@article{wilkinson2016fair,
  author  = {Wilkinson, Mark D. and Dumontier, Michel and Aalbersberg, IJsbrand Jan and Appleton, Gabrielle and Axton, Myles and Baak, Arie and Blomberg, Niklas and Boiten, Jan-Willem and da Silva Santos, Luiz Bonino and Bourne, Philip E. and others},
  title   = {The {FAIR} Guiding Principles for scientific data management and stewardship},
  journal = {Scientific Data},
  volume  = {3},
  pages   = {160018},
  year    = {2016},
  doi     = {10.1038/sdata.2016.18},
  url     = {https://doi.org/10.1038/sdata.2016.18}
}

@article{isayev2017ml-descriptors,
    author={Isayev, Olexandr and Oses, Corey and Toher, Cormac and Gossett, Eric and Curtarolo, Stefano and Tropsha, Alexander},
    title={Universal fragment descriptors for predicting properties of inorganic crystals},
    journal={Nature Communications},
    year={2017},
    month={Jun},
    day={05},
    publisher={The Author(s) SN  -},
    volume={8},
    pages={15679 EP  -},
    note={Article},
    url={http://dx.doi.org/10.1038/ncomms15679}
}

@article{ward2016ml-framework,
    author={Ward, Logan and Agrawal, Ankit and Choudhary, Alok and Wolverton, Christopher},
    title={A general-purpose machine learning framework for predicting properties of inorganic materials},
    journal={Npj Computational Materials},
    year={2016},
    month={Aug},
    day={26},
    publisher={The Author(s) SN  -},
    volume={2},
    pages={16028 EP  -},
    note={Article},
    url={http://dx.doi.org/10.1038/npjcompumats.2016.28}
}

@manual{JSONSchemaDotOrg,
    author = {{JSON Schema}},
    organization = {JSON Schema Project},
    URL= {http://json-schema.org/},
    title = {JSON Schema, official website},
    year = {2026},
}

\end{document}